\documentclass{workshop}
\usepackage{nohyperref}
\begin{document}

\title{
X-ray plateaus in $\gamma$-ray bursts explained by structured jets \\ 
}

\author{
Gor Oganesyan$^{1,}$$^{2,}$$^3$
 \\[12pt]  
$^1$ Gran Sasso Science Institute, Viale F. Crispi 7, I-67100, L’Aquila (AQ), Italy \\
$^2$ INFN - Laboratori Nazionali del Gran Sasso, I-67100, L’Aquila (AQ), Italy \\
$^3$ INAF - Osservatorio Astronomico d’Abruzzo, Via M. Maggini snc, I-64100 Teramo, Italy \\
\textit{E-mail: gor.oganesyan@gssi.it} 
}

\abst{
The follow-up of $\gamma$-ray bursts (GRBs) by the X-ray telescope (XRT, 0.3-10 keV) on board the Neil Gehrels Swift Observatory led to the discovery of a shallow decay phase (the so-called plateau) of the X-ray emission in a good fraction of GRBs. This unexpected temporal behaviour does not fit the standard GRB afterglow expectation. Thus, in the last years, many models emerged, that invoke energy injection into the external shock, requiring long-lasting activity  of the central engine of GRBs. We discuss a new alternative, comprehensive model: the plateau phase originates from the high latitude emission (i.e., the radiation observed from larger angles relative to the line of sight, after the prompt emission from a curved surface is switched off) when the jet exhibits bulk motion and intensity structure. This model enables us to reproduce not only the temporal behaviour of the X-ray light curves, but also the diversity of joint optical-to-X-ray emission.
}

\kword{workshop: proceedings --- gamma-ray bursts}

\maketitle
\thispagestyle{empty}

Before the launch of the Neil Gehrels Swift Observatory (\citet{2004ApJ...611.1005G}, hereafter, {Swift}) in 2004, the observations of $\gamma$-ray burst (GRB) afterglows were performed at relatively late times. The observed light curves were consistent with a basic afterglow theory which suggests a deceleration of the blast wave in the external medium \citep{1998ApJ...497L..17S} and predicts a flux monotonically decaying with time (roughly $\sim$ $t^{-1}$). However, the X-Ray Telescope (XRT, 0.3-10 keV) on board of {Swift} shed new light onto the afterglow physics. Quite often, the soft X-ray afterglows show a plateau phase characterized by a shallow ($\sim$ $t^{-0.5}$ or even flatter) temporal decay segments lasting for thousands of seconds (e.g., \citet{2006ApJ...642..389N}). This feature is inconsistent with a simple afterglow scenario. 

The interpretation of the X-ray plateau is debated and usually requires an energy injection to the forward shock (see \citet{2006ApJ...642..354Z} and references therein). The energy injection into the decelerating forward shock in the circumburst medium for a relatively long time ($ \sim 10^4$ s) is quite challenging and most probably requires the presence of the long-lived, strongly magnetized and highly spinning neutron star as a central engine of a GRB (see \citet{2015JHEAp...7...64B} and references therein). However, the possibility for these neutron stars to form and to provide the required energetics for the GRB jets is quite uncertain. Furthermore, the energy injection models invoked to explain the 
X-ray plateaus fail to interpret commonly observed chromatic behaviour of joint X-ray and optical \citep{2006MNRAS.369..197F}: while the X-ray light curve shows the flat segment, simultaneously observed optical flux is consistent with the standard afterglow scenario. 

Here we briefly discuss a recent model proposed in \citet{2019arXiv190408786O} to explain the X-ray plateau emission. The main idea of this model is based on the out-core jet radiation during the prompt emission (or soon after) when the GRB jet exhibits velocity and the intensity structure. In the uniform jet case, once the impulsive emission is switched off in the entire emitting region, the observer receives gradually a radiation from larger angles from the 
jet axis: later the radiation is received, more de-beamed it is. The so-called high latitude emission \citep{1996ApJ...473..998F,2000ApJ...541L..51K} is widely used to explain the tails of the prompt emission and the initial steep decay phases of the X-ray afterglow (e.g. \citet{2006ApJ...646..351L}). In the structured jet case, the regions outside of the jet core are slower. On the one hand, these regions of the jet are located at higher latitudes, on the second hand, they are less beamed due to the slower bulk motion. This conditions allow for a prolongation of the high latitude emission, provided by the nearly constant Doppler factors. 

\begin{figure}[ht!]
\includegraphics[width=9cm]
{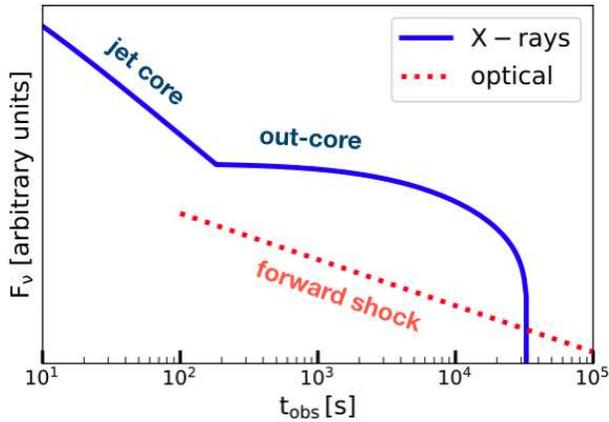}
\caption{Sketch of the chromatic optical and X-ray emissions explained in the out-core jet model. For an on-axis observer, the emission in the optical band is provided from the forward shock, while the X-ray emission is dominated by the high latitude emission from the jet core (initial steep decay) and from the out-core region (plateau and post-plateau phases).}
\label{fig:plateau}
\end{figure}
To illustrate the main advantages of the out-core jet emission model, we 
sketch the idealized chromatic X-ray and optical light curves of GRB afterglows in Fig.\ref{fig:plateau}. In the proposed model, the X-ray emission (blue solid line) is dominated by the high latitude emission from the jet core (the steep decay phase) followed by the plateau and the post-plateau phases arising from the out-core jet regions during (or slightly after) the prompt emission phase. The flux level of the out-core jet emission in the optical bands is much lower than in the soft X-rays since the prompt emission peaks at keV-MeV energy range. Therefore, the optical afterglow emission (red dashed line) is dominated by the radiation from the forward shock. In order to produce the long-lasting plateaus ($\sim 10^4$ s) in the observer's frame, one would require to produce the radiation outside of the jet core while its head size reaches $\sim 10^{15}-10^{16}$ cm. 

The proposed model in \citet{2019arXiv190408786O} is based on the paradigm of the structured jets (e.g., \citet{2001ARep...45..236L,2002MNRAS.332..945R}) which found its wide acceptance after a year of monitoring the multi-wavelength afterglow emission of GRB 170817A (see e.g., \citet{2019Sci...363..968G}). The out-core jet emission during the early phase of GRB is a new way to address the X-ray plateau problem and the broad-band chromatic behaviour of the afterglow emission. Interestingly, the high latitude emission from the jet out-core region is capable to produce two kinds of post-plateau regimes: power-law and/or sharp decays. The latter one is an alternative way out for the so-called internal plateaus sometimes observed in the X-ray light curves (e.g., \citet{2007ApJ...665..599T,2020arXiv200106102S}). 

Given the viewing angle effects and the overlap of the jet out-core emission with the radiation from external shocks, the diverse and intensively observed X-ray afterglows open novel possibilities for the further explorations of the jet structure. The confrontation of the jet-structured based models (see also \citet{2020arXiv200102239B}) with the energy injection model for the X-ray plateau emission is necessary to widen our expectations from the future of the multi-messenger astronomy with the wide field X-ray observatories (e.g., \citet{2018AdSpR..62..191A}) and the next generation gravitational wave detectors (e.g., \citet{2019arXiv191202622M}).

\label{last}

\end{document}